\documentclass[12pt]{article}
\usepackage[a4paper,margin=1.2in]{geometry}

\usepackage[utf8]{inputenc}
\usepackage[T1]{fontenc}    %
\usepackage{hyperref}       %
\usepackage{url}            %
\usepackage{booktabs}       %
\usepackage{amsfonts}       %
\usepackage{amsmath,amssymb}
\usepackage{pifont}%
\usepackage{graphicx}
\usepackage{nicefrac}       %
\usepackage{microtype}      %
\usepackage[acronym, nopostdot, nogroupskip, nohypertypes={acronym,notation}]{glossaries}
\setacronymstyle{long-short}
\usepackage{lipsum}
\usepackage{pythonhighlight}
\usepackage{soul} %
\usepackage{rotating} %
\usepackage{algorithm}
\usepackage{subcaption}

\usepackage{algorithmic}
 \usepackage{setspace}
\let\Algorithm\algorithm
\renewcommand\algorithm[1][]{\Algorithm[#1]\setstretch{1.2}}

\usepackage{tikz}
\usetikzlibrary{arrows.meta,positioning}
\usepackage{xspace}
\usepackage{placeins}
\usepackage{natbib}
\usepackage{mathtools}
\usepackage{wrapfig}

\hypersetup{
    colorlinks=true,
    linkcolor=[rgb]{0.709, 0.070, 0.105},
    citecolor=[rgb]{0.709, 0.070, 0.105},
    filecolor=magenta,      
    urlcolor=[rgb]{0.709, 0.070, 0.105},
}
\newacronym{rkhs}{RKHS}{Reproducing Kernel Hilbert Space}
\newacronym{ipm}{IPM}{integral probability metric}
\newacronym{MCMC}{MCMC}{Markov chain Monte-Carlo}
\newacronym{hmc}{HMC}{Hamiltonian Monte-Carlo}
\newacronym{svi}{SVI}{stochastic variational inference}
\newacronym{SVGD}{SVGD}{Stein Variational Gradient Descent}
\newacronym{vi}{VI}{variational inference}
\newacronym{kl}{KL}{Kullback-Leibler}
\newacronym{iid}{IID}{independent and identically distributed}
\newacronym{ELBO}{ELBO}{evidence lower bound}
\newacronym{RBF}{RBF}{radial basis function}
\newacronym{ARD}{ARD}{Automatic Relevance Determination }
\newacronym{KSD}{KSD}{Kernel Stein Discrepancy}
\newacronym{MMD}{MMD}{Maximum mean discrepancy}
\newacronym{DPP}{DPP}{determinantal point process}
\newacronym{KDPP}{KDPP}{k-determinantal point process}
\newacronym{MAP}{MAP}{maximum a posteriori}
\newacronym{NOx}{NOx}{nitrogen oxide}
\newacronym{UK}{UK}{United Kingdom}

\newacronym{tvd}{TVD}{total variation distance}
\newglossaryentry{GP}
{
  name={GP},
  description={Gaussian process},
  first={\glsentrydesc{GP} (\glsentrytext{GP})},
  plural={GPs},
  descriptionplural={Gaussian processes},
  firstplural={\glsentrydescplural{GP} (\glsentryplural{GP})}
} 
\newglossaryentry{DGP}
{
  name={DGP},
  description={deep Gaussian process},
  first={\glsentrydesc{DGP} (\glsentrytext{DGP})},
  plural={DGPs},
  descriptionplural={deep Gaussian processes},
  firstplural={\glsentrydescplural{DGP} (\glsentryplural{DGP})}
}

\newglossaryentry{RFF}
{
  name={RFF},
  description={Random Fourier Feature},
  first={\glsentrydesc{RFF} (\glsentrytext{RFF})},
  plural={RFF},
  descriptionplural={Random Fourier Features},
  firstplural={\glsentrydescplural{RFF} (\glsentryplural{RFF})}
} 

\usepackage{xcolor}

\definecolor{dkgreen}{rgb}{0,0.6,0}
\definecolor{gray}{rgb}{0.5,0.5,0.5}
\definecolor{mauve}{rgb}{0.58,0,0.82}

\definecolor{midnightblue}{rgb}{0.15, 0.15, 0.5}

\newcommand{\figref}[1]{Figure \ref{#1}}
\newcommand{\secref}[1]{Section \ref{#1}}

\newcommand{\appref}[1]{Appendix \ref{#1}}

\newcommand{\norm}[1]{\left\lVert#1\right\rVert}

\newcommand{\x}{\mathbf{x}}

\newcommand{\by}{\mathbf{y}}

\newcommand{\GP}{\mathcal{GP}}

\newcommand{\notwo}{NO\textsubscript{2}\xspace}

\newcommand{\bu}{\mathbf u}

\newcommand{\f}{\mathbf{f}}
\newcommand{\given}{\,|\,}
\newcommand{\btheta}{\boldsymbol\theta}

\newcommand{\bparticle}{\boldsymbol\lambda}

\newcommand{\cO}{\mathcal{O}}
\newcommand{\cD}{\mathcal{D}}

\title{\textbf{A Probabilistic Assessment of the COVID-19 Lockdown on Air Quality in the UK}}
\author{Thomas Pinder$^{1}$, Michael Hollaway$^{2}$, Christopher Nemeth$^1$ \\
Paul J. Young$^{3}$, David Leslie$^1$ \\ \vspace{-0.2cm}
{\footnotesize $^1$Department of Mathematics and Statistics, Lancaster University, UK} \\ \vspace{-0.2cm}
{\footnotesize $^2$UK Centre for Ecology \& Hydrology, Lancaster Environment Centre, Lancaster University, UK} \\
{\footnotesize $^3$Lancaster Environment Centre, Lancaster University, UK }
}
\date{\today}

\begin{document}

\maketitle

\begin{abstract}
In March 2020 the \gls{UK} entered a nationwide lockdown period due to the Covid-19 pandemic. As a result, levels of \notwo in the atmosphere dropped. In this work, we use over 550,134 \notwo data points from 237 stations in the \gls{UK} to build a spatiotemporal \gls{GP} capable of predicting \notwo levels across the entire \gls{UK}. We integrate several covariate datasets to enhance the model's ability to capture the complex spatiotemporal dynamics of \notwo. Our numerical analyses show that, within two weeks of a \gls{UK} lockdown being imposed, UK \notwo levels dropped 36.8\%. Further, we show that as a direct result of lockdown \notwo levels were 29-38\% lower than what they would have been had no lockdown occurred. In accompaniment to these numerical results, we provide a software framework that allows practitioners to easily and efficiently fit similar models.
\end{abstract}

\section{Introduction}

In response to the COVID-19 pandemic, many countries resorted to ordering citizens to remain at home. These lockdown measures had the effect of curtailing certain polluting activities, resulting in reduced pollutant concentrations in many places around the world \citep[e.g.][]{forster_current_2020}. Clear evidence of reductions has been found in nitrogen dioxide (\notwo), a pollutant that can negatively impact respiratory health \citep{henschel_health_2013,anenberg_estimates_2018} and which is strongly associated with transport emissions due to the burning of fossil fuels \citep{sundvor_road_nodate}. Yet these (often provisional) assessments of lockdown impacts on \notwo have largely been restricted to analyses of individual measurement locations or satellite retrievals and focused on daily means \citep{berman_changes_2020, carslaw_covid-19_2020, schindler_svs_2020}, rather than combining information across different sources. In this paper we utilise a spatiotemporal approach to model the patterns of change in \notwo.

Modelling continuous spatiotemporal surfaces using a set of discrete observations is a common task in geostatistics \citep{diggle_model-based_2007}. 
This approach assumes that there exists a latent partially observed function which is commonly modelled with a \gls{GP}. Consequently, it is then possible to construct a fully-Bayesian model. Historically these models have been computationally challenging due to their $\cO(n^3)$ cost, where $n$ is the number of datapoints. However, recent advances in the sparse \gls{GP} literature has reduced this cost to $\cO(nm^2)$, where $m<<n$ \citep{hensman_gaussian_2013}.

In this work we combine several measurement datasets and covariate information to build a highly informative spatiotemporal model that yields novel insights into the affect of the COVID-19 \gls{UK} lockdown on \notwo levels. To build a model that can be both computationally efficient to fit and powerful enough to capture the complex latent process, we use recent developments in the \gls{GP} literature by using the SteinGP presented by \cite{pinder_stein_2020}. Finally, we examine the inferred latent function to report two quantitative results concerning the change in \notwo levels. The first gives the integrated \notwo levels pre and post-lockdown, and the second compares the latent function post-lockdown to the \notwo levels that, accounting for weather-driven variability, would have been recorded had a lockdown not been instigated. 

\section{Background}

\subsection{Environmental datasets}\label{sec:background:data}
\paragraph{Ground measurements} Hourly \notwo concentrations are recorded by the Automatic Urban and Rural Network (AURN) that covers England and Northern Ireland, the Scottish Air Quality Network (SAQN) and the Welsh Air Quality Network (WAQN) groups of sensors. This network is comprised of 237 sites (shown in \appref{app:stationLocation}) that are used to monitor compliance of UK air quality with ambient air quality directives. It provides hourly concentrations of a number of key pollutants, including \notwo. The data covers a wide range of site types from urban roadside through to rural settings\footnote{https://uk-air.defra.gov.uk/networks/network-info?view=aurn}. The \notwo values are pre-processed by the RMWeather package \cite{grange_random_2018} to normalise for the effects of weather, namely wind speed and wind direction.

\paragraph{Land cover data} The UK Centre for Ecology and Hydrology Land cover Map (LCM) 2019 describes the land cover over Great Britain \citep{Mortonetal2020a} and Northern Ireland \citep{Mortonetal2020b} at 25\,m resolution for 21 broad classes. Here, the 25\,m data is aggregated to 200\,m grid squares over the UK and the proportion of each square that falls under one of the 3 urban classes is calculated. This gives an estimate of the urban fraction in each grid cell.

\begin{wrapfigure}{l}{0.48\textwidth}
  \begin{center}
    \includegraphics[width=0.47\textwidth]{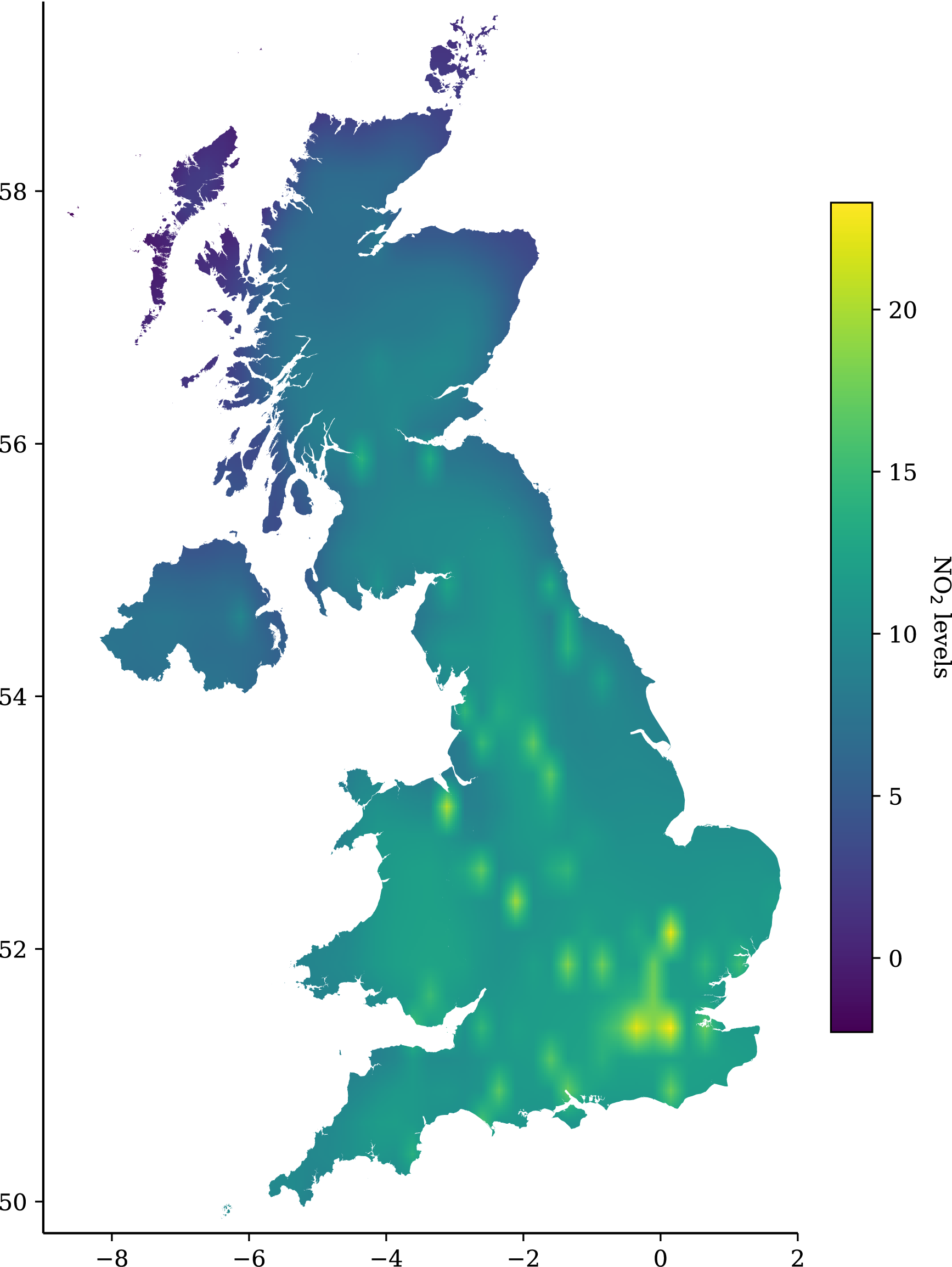}
  \end{center}
  \caption{Inferred \notwo spatial surface ($\mu$gm$^{-3}$) in the \gls{UK} at 9AM on March 23$^{\text{rd}}$; the day that initial lockdown measures were announced.}
  \label{fig:March16thScrengrab}
  \vspace{-0.5cm}
\end{wrapfigure}

\paragraph{Reanalysis data} ERA5 is a reanalysis product developed by the European Centre for Medium Range Weather Forecasting (ECMWF). Using data assimilation techniques with a forecast model, satellite data and ground-based observations, global hourly meteorological conditions are produced at a horizontal resolution of $\approx$31\,km \citep{hersbach_era5_2020}, from 1979 through to the present day. Here wind speed, wind direction, relative humidity and temperature data are used to provide our model with meteorological covariate information \citep{grange_random_2018}. 

In the model described in \secref{sec:modelSpecific} we will refer to the land cover and reanalysis data as \textit{covariates}, the location of the ground measurement stations as the \textit{spatial} dimensions and the timestamp as the \textit{temporal} dimension. This selection of covariates is consistent with other air pollution studies \citep[e.g.,][]{grange_random_2018}.

\subsection{Stein variational Gaussian processes}\label{sec:background:gp}
Considering a set of $n$ observations\footnote{We use capitalised, lowercase bold and regular lowercase to denote matrices, vectors and scalars, respectively.} $\mathcal{D}=\{X, \by\}=\{\mathbf{x}_i, y_i\}_{i=1}^n$ where $\mathbf{x}_i \in \mathbb{R}^n$ and $y_i \in \mathbb{R}$. We seek to learn a function $f: \mathbb{R}^d\rightarrow\mathbb{R}$ that relates inputs to outputs given $\cD$ and a likelihood function $p(\by \given f)=\mathcal{N}(\f, \sigma_n^2)$ where $\f=f(X)$ is the realisation of $f$ at the input locations $X$. We place a \gls{GP} prior on $f$ such that $p(f) \sim \GP(\mu, k_{\btheta})$, characterised by a mean function $\mu: \mathcal{X}\rightarrow\mathbb{R}$ and a $\btheta$-parameterised kernel function $k_{\btheta}: \mathcal{X}\times\mathcal{X}\rightarrow\mathbb{R}$.

To introduce sparsity into our \gls{GP} model we introduce a set of \textit{inducing points} $Z$ that are used to approximate $X$. This is common practice for \gls{GP} modelling with large datasets as it enables tractable computation \citep[see][]{snelson_sparse_2006, hensman_gaussian_2013}. Using recent work by \cite{pinder_stein_2020}, the kernel parameters $\btheta$, observational noise $\sigma^2$ and latent values $\bu = f(Z)$ of the \gls{GP} can be learned using \gls{SVGD}. Such an approach allows for highly efficient computations to be done, whilst allowing the practitioner to place prior distributions on parameters, a quality not enjoyed by regular sparse frameworks. Further, using the software package described by \cite{pinder_stein_2020}, we demonstrate how practitioners can easily integrate these models into their existing analyses. See \appref{app:demoImp} for an example of this.

\section{Analysis}\label{sec:analysis}

\subsection{Model design}\label{sec:modelSpecific}

\begin{figure}
  \centering
    \includegraphics[width=0.8\textwidth]{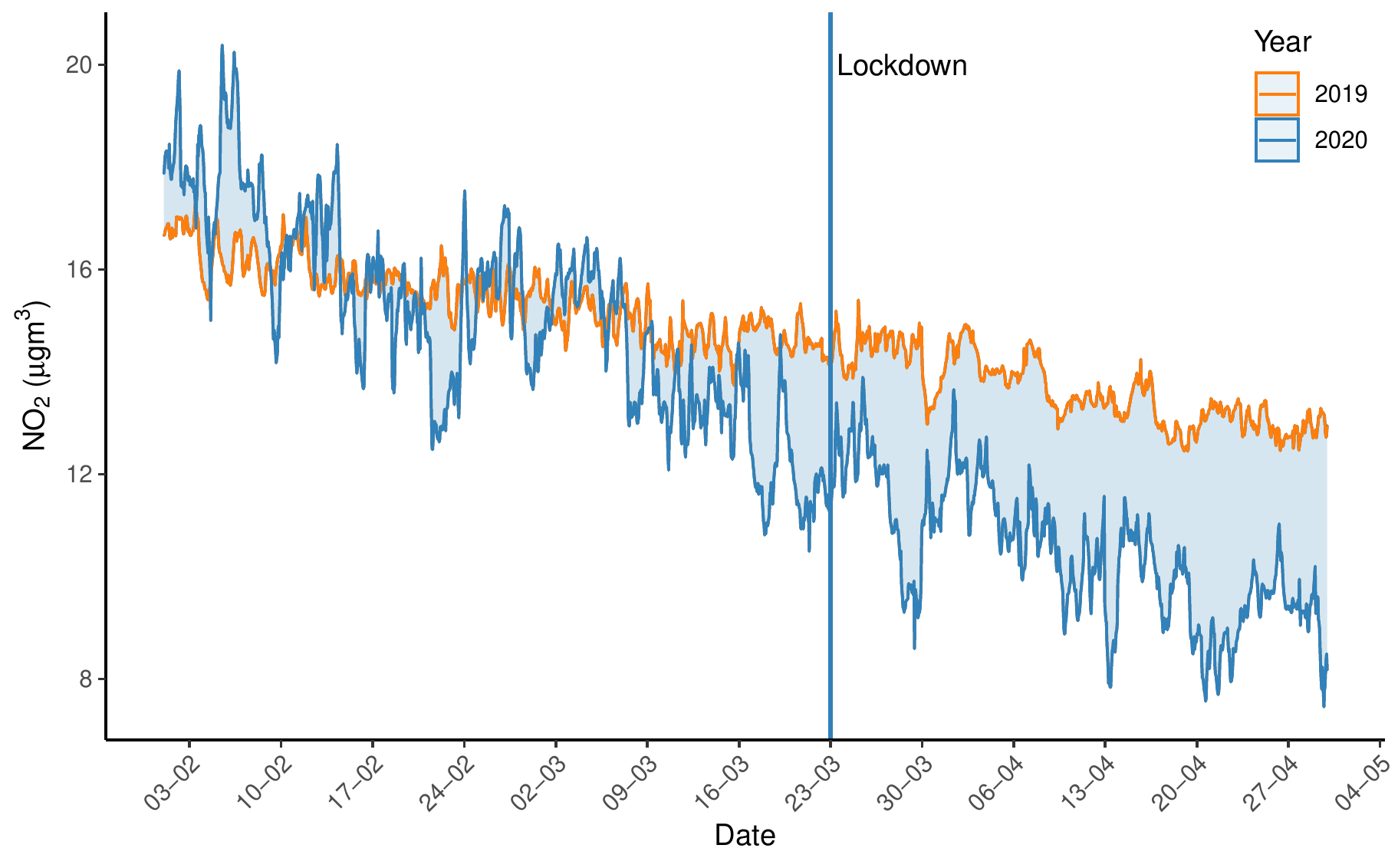}
  \caption{The difference in the inferred latent trend in 2019 and 2020 in London, United Kingdom. The start of the UK lockdown on March 23$^{rd}$ is depicted by a blue vertical line.}
  \label{fig:YearlyTrends}
\end{figure}

We formulate the following kernel\footnote{For notational conciseness, we drop individual kernel's dependence on parameters and use $\btheta$ to denote the union of parameters from all kernels.} to capture the complex spatiotemporal dynamics of \notwo, 
\begin{align}
    \label{equn:modelKernel}
    k_{\btheta}(\mathbf{x}, \mathbf{x}') = k_1(\mathbf{x}_{s},\mathbf{x}_{s}')k_2(\mathbf{x}_{t},\mathbf{x}_{t}')k_3(\mathbf{x}_{c},\mathbf{x}_{c}')
\end{align}
where the subscripts $s$, $t$ and $c$ denote the spatial, temporal and covariate dimensions of the data. A third-order Mat\'ern kernel is used for the spatial kernel and a squared exponential kernel for the covariate values. Furthermore, we use a product kernel of a non-stationary third-order polynomial kernel and first-order Mat\'ern kernel for the temporal dimension. A white noise process is applied to all seven input dimensions. Full kernel expressions can be found in \appref{sec:app:kernelDesign:expressions}.

The choice of kernels was driven by two factors: firstly our beliefs about the spatiotemporal behaviour of \notwo, and secondly from a comparison of the performance of several different kernels on a held-out set of data (see \appref{sec:app:kernelDesign:alt}).

To supplement the above kernel, we also equip the \gls{GP} with a linear mean function $y_i = \mathbf{a}^{\top} \x_i + b$ where $\mathbf{a} \in \mathbb{R}^{d}$ and $b \in \mathbb{R}$. The rationale for including a linear mean function is that if we focus on the temporal dimension, there is a globally decreasing trend in \notwo levels. Including a linear mean function provides an effective method to capture this.

\subsection{Inference}

From the above model design, our full set of parameters becomes $\Omega = \{\btheta, \mathbf{a}, b, \sigma^2, \bu\}$ where $|\Omega|=1022$\footnote{As per the expressions in \appref{tab:kernels}, the model contains 13 kernel hyperparameters, 8 mean function hyperparameters the observational noise $\sigma^2$ and 1000 latent values $\bu$.} The learned values of these parameters are crucial as it will govern the behaviour of the inferred latent function. Within $\btheta$ we have three types of parameter: a lengthscale $\ell$ that will control the amount of correlation between datapoints, a variance $\alpha$ that controls the magnitude of the latent function and polynomial coefficients $\gamma$ that control the temporal behaviour of the \gls{GP}.%
To learn posterior distributions for each of these parameters, we independently initialise $J=10$ particles to be used within the \gls{SVGD} optimisation update step. We employ the Adam \citep{kingma_adam:_2015} optimiser to estimate step-sizes $\epsilon$ and compute the gradients using mini-batches containing 250 observations.

We fit the model from \secref{sec:modelSpecific} to 89 days worth of hourly data from February 1$^{\text{st}}$ to April 30$^{\text{th}}$, noting that lockdown measures were first imposed in the \gls{UK} on March 23$^{\text{rd}}$. It is of note that enhanced social distancing was introduced in the UK from March 16$^{\text{th}}$. A snapshot of the inferred spatial surface on March 23$^{\text{rd}}$ at 9AM can be seen in \figref{fig:March16thScrengrab}.

A set of 1000 inducing points $Z$ are initialised using a \gls{KDPP} \citep{burt_convergence_2020} where $k=1000$ with 10000 samples drawn according to the \gls{MCMC} sampler described in \cite{anari_monte_2016}. As is standard, we treat the latent values $\bu = f(Z)$ as a model parameter.

\section{Conclusions}

\paragraph{\notwo reduction} From the inferred latent process, we are able go beyond assessing changes in \notwo levels at the location of measurement stations. Using the latent process we can instead compute integrated spatiotemporal mean levels. This is desirable as it gives quantities that are representative of the entire \gls{UK}, and not just the parts of the \gls{UK} where measurement stations exist. For the entire \gls{UK}, the integrated spatial \notwo levels decreased 36.8\% between 9AM on March 16$^{\text{th}}$ and 9AM on March 30$^{\text{th}}$ (one week pre and post-lockdown).

\paragraph{Deviation from \textit{normal} trends}While a clear reduction in \notwo is apparent across the pre and post lockdown period, we would also like to give more context to the apparent decrease by comparing to levels in previous years. To define a \textit{normal} trend, we consider data from February 1$^{\text{st}}$ to April 30$^{\text{th}}$ in 2019. Comparing the latent function that is learned in 2019 to the latent function that corresponds to data from the same period in 2020, we can see the effect that lockdown had on \notwo levels, relative to expected temporal changes from 2019.

\begin{wrapfigure}{r}{0.5\textwidth}
  \begin{center}
    \includegraphics[width=0.5\textwidth]{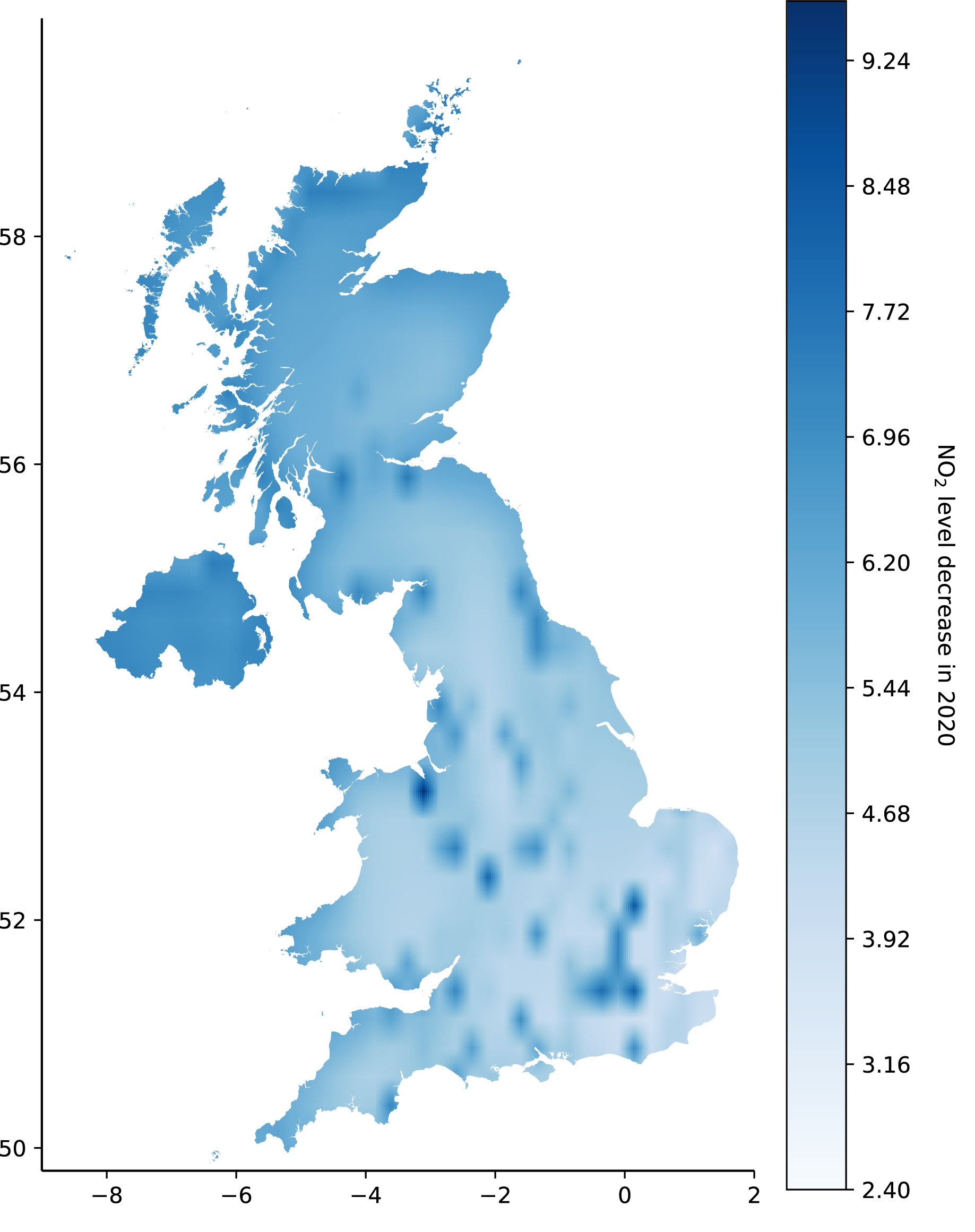}
  \end{center}
  \caption{Difference in spatial means ($\mu$gm$^{-3}$) integrated over March 23$^{\text{rd}}$ to April 20$^{\text{th}}$ in 2019 and 2020. Darker shades of blue indicate a larger decrease from 2019 to 2020.}
  \label{fig:SptialMeans}
  \vspace{-0.5cm}
\end{wrapfigure}

Quantitatively, by computing the integrated spatiotemporal mean for data from February 1$^{\text{st}}$ through to the March 23$^{\text{rd}}$ in 2019 and 2020, we can see that \notwo levels are 9\% lower in 2020 compared to levels in 2019. However, if we now compute the integrated spatiotemporal mean for data from March 23$^{\text{rd}}$ through to April 30$^{\text{th}}$ in 2019 and 2020, then we can see that levels in 2020 are now 38\% lower. This can be seen in \figref{fig:SptialMeans} where we plot the difference in \notwo levels from 2019 to 2020, integrated over the temporal values from March 23$^{\text{rd}}$ to April 30$^{\text{th}}$. In all locations \notwo values are lower in 2020 when compared to 2019. Moreover, we can see that the decrease is greater in cities and the more urban areas of the UK compared to the more rural regions. For illustrative purposes, we can also see this temporal change in \figref{fig:YearlyTrends}. Here we observe the inferred time series at a single location (51.5$^o$N, 0.1$^o$W) that corresponds to Westminster in London, UK, where it is apparent that the divergence in \notwo levels begins around March 23$^{\text{rd}}$.

\paragraph{Future directions}This work has demonstrated a method to infer spatially and temporally complete air quality measurement data, which will serve machine learners, environmental scientists and policymakers. Computing spatiotemporal interpolations in the way described in this article allows us to make informed decisions based upon air quality levels in locations where we have no sensors. Further, using the associated predictive uncertainty, we can express to policy makers the \textit{confidence} we have in these interpolations. Overall, this is a very powerful technique for atmospheric and environmental scientists. 

Whilst we have focussed our analysis on the COVID-19 lockdown in the \gls{UK}, there is nothing preventing this analysis being applied to different time periods, locations and environmental variables. For instance, this could be used in conjunction with UK-wide health data to identify regions of the \gls{UK} most at risk of air quality related illnesses. Additionally, due to the accompanying uncertainty estimates in the predictive output of a \gls{GP}, there is scope for this work to be used in a decision making framework to decide where new sensors should be placed. Perhaps the simplest metric for informing this decision is choosing the point location that results in the largest decrease in predictive uncertainty. Finally, this work could also provide a dataset suitable for understanding and quantifying the spatiotemporal evolution of pollution - or indeed any sparsely measured environmental variable - and for evaluating process models of the atmosphere and climate.

\clearpage

\bibliographystyle{apalike}
\bibliography{main}

\begin{thebibliography}{}

\bibitem[Abadi et~al., 2016]{abadi_tensorflow_2016}
Abadi, M., Agarwal, A., Barham, P., Brevdo, E., Chen, Z., Citro, C., Corrado,
  G.~S., Davis, A., Dean, J., Devin, M., Ghemawat, S., Goodfellow, I., Harp,
  A., Irving, G., Isard, M., Jia, Y., Jozefowicz, R., Kaiser, L., Kudlur, M.,
  Levenberg, J., Mane, D., Monga, R., Moore, S., Murray, D., Olah, C.,
  Schuster, M., Shlens, J., Steiner, B., Sutskever, I., Talwar, K., Tucker, P.,
  Vanhoucke, V., Vasudevan, V., Viegas, F., Vinyals, O., Warden, P.,
  Wattenberg, M., Wicke, M., Yu, Y., and Zheng, X. (2016).
\newblock {TensorFlow}: {Large}-{Scale} {Machine} {Learning} on {Heterogeneous}
  {Distributed} {Systems}.
\newblock {\em arXiv:1603.04467 [cs]}.
\newblock arXiv: 1603.04467.

\bibitem[Anari et~al., 2016]{anari_monte_2016}
Anari, N., Gharan, S.~O., and Rezaei, A. (2016).
\newblock Monte carlo markov chain algorithms for sampling strongly rayleigh
  distributions and determinantal point processes.
\newblock In Feldman, V., Rakhlin, A., and Shamir, O., editors, {\em
  Proceedings of the 29th Conference on Learning Theory, {COLT} 2016, New York,
  USA, June 23-26, 2016}, volume~49 of {\em {JMLR} Workshop and Conference
  Proceedings}, pages 103--115. JMLR.org.

\bibitem[Anenberg et~al., 2018]{anenberg_estimates_2018}
Anenberg, S.~C., Henze, D.~K., Tinney, V., Kinney, P.~L., Raich, W., Fann, N.,
  Malley, C.~S., Roman, H., Lamsal, L., Duncan, B., Martin, R.~V., van
  Donkelaar, A., Brauer, M., Doherty, R., Jonson, J.~E., Davila, Y., Sudo, K.,
  and Kuylenstierna, J.~C. (2018).
\newblock Estimates of the {Global} {Burden} of {Ambient} {PM2}.5, {Ozone}, and
  {NO2} on {Asthma} {Incidence} and {Emergency} {Room} {Visits}.
\newblock {\em Environmental Health Perspectives}, 126(10):107004.

\bibitem[Berman and Ebisu, 2020]{berman_changes_2020}
Berman, J.~D. and Ebisu, K. (2020).
\newblock Changes in {U}.{S}. air pollution during the {COVID}-19 pandemic.
\newblock {\em Science of The Total Environment}, 739:139864.

\bibitem[Burt et~al., 2020]{burt_convergence_2020}
Burt, D.~R., Rasmussen, C.~E., and van~der Wilk, M. (2020).
\newblock Convergence of {Sparse} {Variational} {Inference} in {Gaussian}
  {Processes} {Regression}.
\newblock {\em arXiv:2008.00323 [cs, stat]}.
\newblock arXiv: 2008.00323.

\bibitem[Carslaw, 2020]{carslaw_covid-19_2020}
Carslaw, D. (2020).
\newblock {COVID}-19 and changes in air pollution.
\newblock Technical report.

\bibitem[Carslaw and Ropkins, 2012]{carslaw_openair_2012}
Carslaw, D.~C. and Ropkins, K. (2012).
\newblock openair — {An} {R} package for air quality data analysis.
\newblock {\em Environmental Modelling \& Software}, 27-28:52--61.

\bibitem[Diggle and Ribeiro, 2007]{diggle_model-based_2007}
Diggle, P. and Ribeiro, P.~J. (2007).
\newblock {\em Model-based {Geostatistics}}.
\newblock Springer {Series} in {Statistics}. Springer-Verlag, New York.

\bibitem[Forster et~al., 2020]{forster_current_2020}
Forster, P.~M., Forster, H.~I., Evans, M.~J., Gidden, M.~J., Jones, C.~D.,
  Keller, C.~A., Lamboll, R.~D., Quéré, C.~L., Rogelj, J., Rosen, D.,
  Schleussner, C.-F., Richardson, T.~B., Smith, C.~J., and Turnock, S.~T.
  (2020).
\newblock Current and future global climate impacts resulting from {COVID}-19.
\newblock {\em Nature Climate Change}, pages 1--7.
\newblock Publisher: Nature Publishing Group.

\bibitem[Grange et~al., 2018]{grange_random_2018}
Grange, S.~K., Carslaw, D.~C., Lewis, A.~C., Boleti, E., and Hueglin, C.
  (2018).
\newblock Random forest meteorological normalisation models for {Swiss}
  {PM}$_{\textrm{10}}$ trend analysis.
\newblock {\em Atmospheric Chemistry and Physics}, 18(9):6223--6239.
\newblock Publisher: Copernicus GmbH.

\bibitem[Henschel and Chan, 2013]{henschel_health_2013}
Henschel, S. and Chan, G. (2013).
\newblock Health risks of air pollution in {Europe} – {HRAPIE} project. {New}
  emerging risks to health from air pollution –results from the survey of
  experts.
\newblock Technical report, World Health Organisation.

\bibitem[Hensman et~al., 2013]{hensman_gaussian_2013}
Hensman, J., Fusi, N., and Lawrence, N.~D. (2013).
\newblock Gaussian processes for big data.
\newblock In Nicholson, A. and Smyth, P., editors, {\em Proceedings of the
  Twenty-Ninth Conference on Uncertainty in Artificial Intelligence, {UAI}
  2013, Bellevue, WA, USA, August 11-15, 2013}. {AUAI} Press.

\bibitem[Hersbach et~al., 2020]{hersbach_era5_2020}
Hersbach, H., Bell, B., Berrisford, P., Hirahara, S., Horányi, A.,
  Muñoz-Sabater, J., Nicolas, J., Peubey, C., Radu, R., Schepers, D., Simmons,
  A., Soci, C., Abdalla, S., Abellan, X., Balsamo, G., Bechtold, P., Biavati,
  G., Bidlot, J., Bonavita, M., De~Chiara, G., Dahlgren, P., Dee, D.,
  Diamantakis, M., Dragani, R., Flemming, J., Forbes, R., Fuentes, M., Geer,
  A., Haimberger, L., Healy, S., Hogan, R.~J., Hólm, E., Janisková, M.,
  Keeley, S., Laloyaux, P., Lopez, P., Lupu, C., Radnoti, G., de~Rosnay, P.,
  Rozum, I., Vamborg, F., Villaume, S., and Thépaut, J.-N. (2020).
\newblock The {ERA5} global reanalysis.
\newblock {\em Quarterly Journal of the Royal Meteorological Society},
  146(730):1999--2049.
\newblock tex.eprint:
  https://rmets.onlinelibrary.wiley.com/doi/pdf/10.1002/qj.3803.

\bibitem[Kingma and Ba, 2015]{kingma_adam:_2015}
Kingma, D.~P. and Ba, J. (2015).
\newblock Adam: {A} method for stochastic optimization.
\newblock In Bengio, Y. and LeCun, Y., editors, {\em 3rd International
  Conference on Learning Representations, {ICLR} 2015, San Diego, CA, USA, May
  7-9, 2015, Conference Track Proceedings}.

\bibitem[Liu and Wang, 2016]{liu_stein_2016}
Liu, Q. and Wang, D. (2016).
\newblock Stein variational gradient descent: {A} general purpose bayesian
  inference algorithm.
\newblock In Lee, D.~D., Sugiyama, M., von Luxburg, U., Guyon, I., and Garnett,
  R., editors, {\em Advances in Neural Information Processing Systems 29:
  Annual Conference on Neural Information Processing Systems 2016, December
  5-10, 2016, Barcelona, Spain}, pages 2370--2378.

\bibitem[Matthews et~al., 2017]{matthews_gpflow:_2017}
Matthews, A. G. d.~G., Wilk, M. v.~d., Nickson, T., Fujii, K., Boukouvalas, A.,
  Le\{{\textbackslash}'o\}n-Villagr\{{\textbackslash}'a\}, P., Ghahramani, Z.,
  and Hensman, J. (2017).
\newblock {GPflow}: {A} {Gaussian} {Process} {Library} using {TensorFlow}.
\newblock {\em Journal of Machine Learning Research}, 18(40):1--6.

\bibitem[Morton et~al., 2020a]{Mortonetal2020b}
Morton, R.~D., Marston, C.~G., O'Neil, A.~W., and Rowland, C.~S. (2020a).
\newblock Land cover map 2019 (25m rasterised land parcels, n. {Ireland}).
\newblock Publisher: NERC Environmental Information Data Centre.

\bibitem[Morton et~al., 2020b]{Mortonetal2020a}
Morton, R.~D., Marston, C.~G., O’Neil, A.~W., and Rowland, C.~S. (2020b).
\newblock Land {Cover} {Map} 2019 (25m rasterised land parcels, {GB}).
\newblock Publisher: NERC Environmental Information Data Centre.

\bibitem[Pinder et~al., 2020]{pinder_stein_2020}
Pinder, T., Nemeth, C., and Leslie, D. (2020).
\newblock Stein {Variational} {Gaussian} {Processes}.
\newblock {\em arXiv:2009.12141 [cs, stat]}.
\newblock arXiv: 2009.12141.

\bibitem[Schindler, 2020]{schindler_svs_2020}
Schindler, T. (2020).
\newblock {SVS}: {Reductions} in {Pollution} {Associated} with {Decreased}
  {Fossil} {Fuel} {Use} {Resulting} from {COVID}-19 {Mitigation}.
\newblock Technical report.

\bibitem[Snelson and Ghahramani, 2005]{snelson_sparse_2006}
Snelson, E. and Ghahramani, Z. (2005).
\newblock Sparse gaussian processes using pseudo-inputs.
\newblock In {\em Advances in Neural Information Processing Systems 18 [Neural
  Information Processing Systems, {NIPS} 2005, December 5-8, 2005, Vancouver,
  British Columbia, Canada]}, pages 1257--1264.

\bibitem[Sundvor et~al., 2013]{sundvor_road_nodate}
Sundvor, I., Nuria, C.~B., Viana, M., Querol, X., Reche, C., Amato, F.,
  Mellios, G., and Guerreiro, C. (2013).
\newblock Road traffic’s contribution to air quality in {European} cities
  {ETC}/{ACM} {Technical} {Paper} 2012/14.
\newblock Technical report.

\end{thebibliography}
\clearpage

\appendix
\section{Monitoring stations}\label{app:stationLocation}

\begin{figure}[ht]
    \centering
    \includegraphics[width=0.9\textwidth]{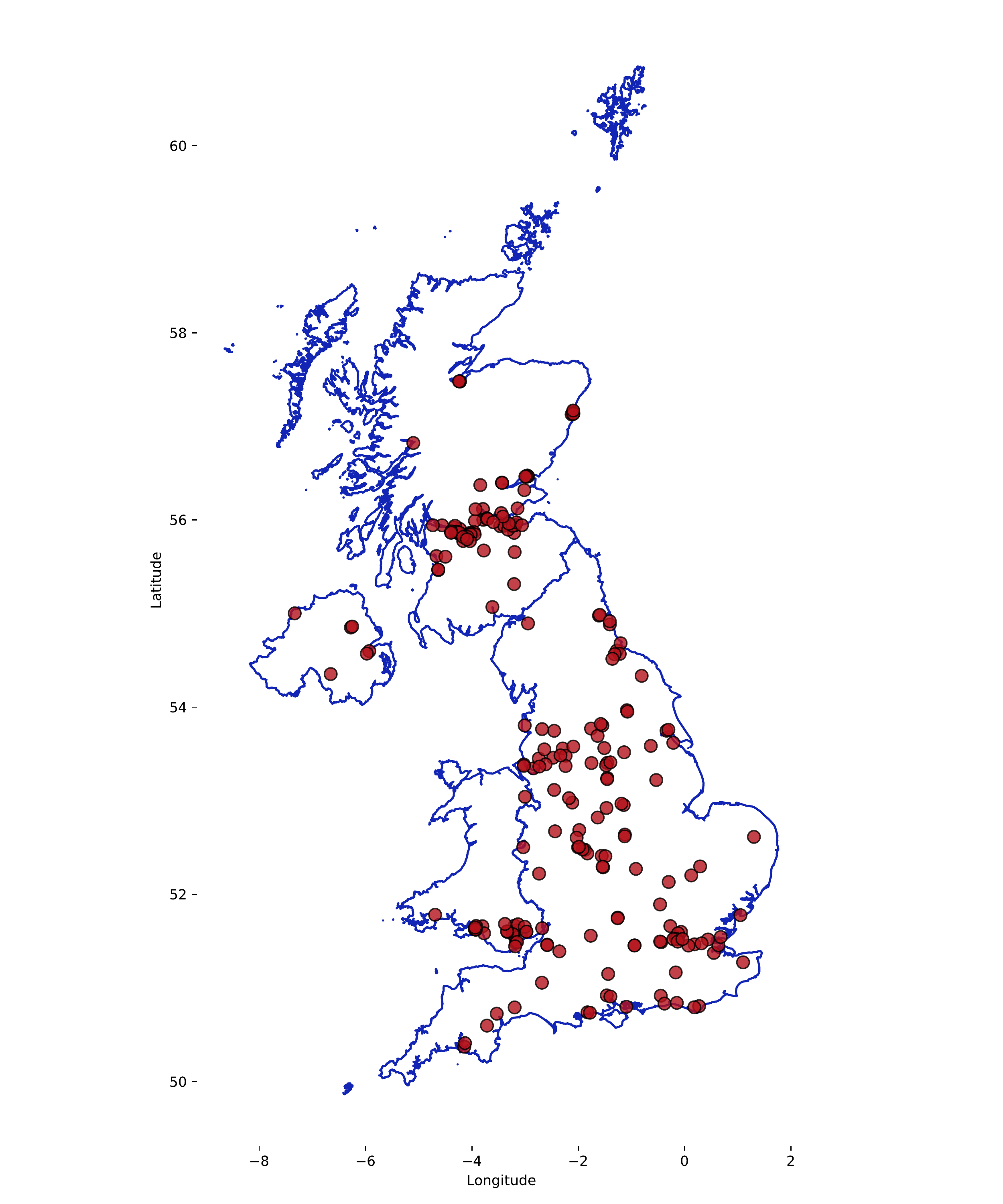}
    \caption{Spatial dispersion of the AURN, SAQN and WAQN \notwo measurement stations described in \secref{sec:background:data}.}
    \label{fig:stationLocation}
\end{figure}

\section{Kernel design}\label{sec:app:kernelDesign}

\subsection{Kernels}\label{sec:app:kernelDesign:expressions}

\begin{table}[!ht]
\centering
\caption{Explicit forms of the kernels used in \secref{sec:analysis} operating on an arbitrary $\x \in \mathbb{R}^d$. For Mat\'ern kernels, the respective order is given by $\nicefrac{c}{2}-2$. For notational brevity we let $\tau = \norm{\x, \x'}_2^2$.}
\label{tab:kernels}
\begin{tabular}{l|c|c}
Kernel & $k_{\btheta}(\x, \x')$ & $\btheta$  \\ \hline
Mat\'ern \nicefrac{1}{2} & $\sigma^2\exp(\nicefrac{-\tau}{\boldsymbol{\ell}})$ & $\{\sigma \in \mathbb{R}, \boldsymbol{\ell}\in \mathbb{R}^d\}$   \\
Mat\'ern \nicefrac{5}{2} & $\sigma^2(1+\nicefrac{\sqrt{5}\tau}{\boldsymbol{\ell}}+\nicefrac{\nicefrac{5}{3}\tau^2}{\boldsymbol{\ell}^2})\exp(-\nicefrac{\sqrt{5}\tau}{\boldsymbol{\ell}})$ & $\{\sigma \in \mathbb{R}, \boldsymbol{\ell}\in \mathbb{R}^d\}$   \\
Squared exponential & $\sigma^2\exp(\nicefrac{-\tau}{2\boldsymbol{\ell}^2})$ & $\{\sigma \in \mathbb{R}, \boldsymbol{\ell}\in \mathbb{R}^d\}$   \\
Polynomial ($d$-order)  & $(\sigma^2\x^{\top}\x' + \gamma)^{d}$ & $\{\sigma \in \mathbb{R}, \gamma \in \mathbb{R}\}$ \\ 
White &     $\begin{cases}
      \sigma & \text{if $\x = \x'$}\\
      0 & \text{otherwise}
    \end{cases}$   & $\{\sigma \in \mathbb{R}\}$
    \end{tabular}
\end{table}

\subsection{Alternative kernels}\label{sec:app:kernelDesign:alt}
In the course of building the \gls{GP} model described in \secref{sec:analysis}, several alternative models were considered. Primarily, this is driven by the use of increasingly complex kernels whilst trying to build the most parsimonious model. Where complexity is determined by the number of parameters, we will proceed to detail the alternative kernels that were considered and their respective metrics in order of increasing complexity. 1) isotropic third-order Mat\'ern, 2) \gls{ARD} third-order Mat\'ern, 3) third-order polynomial kernel acting on the data's temporal dimension convolved with an \gls{ARD} third-order Mat\'ern acting on all but the temporal dimension. In each case a zero and linear mean function are considered. 

As the kernel's complexity increased, the the marginal log-likelihood increased and the root mean-squared error on a held-out set of data decreased. Based upon these two metrics, the kernel described in \secref{sec:analysis} was used.

\FloatBarrier

\FloatBarrier

\section{Stein variational Gaussian processes}\label{app:SteinGP}
Building upon the \gls{GP} setup in \secref{sec:background:gp}, we give a brief description of Stein variational Gaussian processes for the interested reader. 

Following \cite{pinder_stein_2020}, we introduce a set of $J$ particles $\{\bparticle^i\}_{i=1}^J$ where $\bparticle_i = \{\btheta, \sigma_n^2, \bu\}$. The particles are then jointly optimised according to the mapping $\mathcal{T}(\bparticle_{t})=\bparticle_t + \epsilon_t \zeta(\bparticle_t)$ where $\mathcal{T}$ is the velocity field that yields the fastest reduction of the Kullback-Leibler divergence from the approximate posterior to the true posterior. Further, $\epsilon$ is a step-size parameter. Empirically, this velocity field can be computed by 
\begin{align}
    \label{equn:SVGD_update}
    \hat{\zeta}(\bparticle) = \frac{1}{J}\sum^J_{j=1}\bigg[\underbrace{\kappa(\bparticle^{j}_{t}, \bparticle)\nabla_{\bparticle}%
    \log p(\bparticle_{t}^{j})}_{\text{Attraction}} + \underbrace{\nabla_{\bparticle}%
    \kappa(\bparticle_{t}^{j}, \bparticle)}_{\text{Repulsion}}  \bigg].
\end{align}
where $\kappa: \mathcal{X} \times \mathcal{X} \rightarrow \mathbb{R}$ is a kernel function, such as the \gls{RBF} $\kappa(\mathbf{x}, \mathbf{x'}) = \sigma^2\exp(\nicefrac{-\norm{\mathbf{x}-\mathbf{x}'}_2^2}{2\ell^{2}})$. The first term in \eqref{equn:SVGD_update} attracts particles to areas of high probability mass in Gaussian process posterior, whereas the second term in \eqref{equn:SVGD_update} encourages diversity amongst particles. For more details, see \cite{liu_stein_2016, pinder_stein_2020}.

\section{Training details}\label{sec:app:expDetails}

\paragraph{Prior distributions} For the lengthscale parameters present in stationary kernels we use a Gamma prior with shape and scale parameters of 1. All variance parameters are assigned a Gamma prior with shape and scale parameters of 2. For the mean function's coefficients $\mathbf{a}$  and intercept we assign a unit Gaussian prior.

\paragraph{SVGD kernels} In our implementation of \gls{KSD} the \gls{RBF} kernel is used to compute \eqref{equn:SVGD_update}. The kernel's lengthscale is estimated at each iteration of the optimisation procedure using the median rule, as per \cite{liu_stein_2016}.

\paragraph{Inducing points} The inducing points used in \secref{sec:analysis} are set using a \gls{KDPP} as per \cite{burt_convergence_2020}.

\paragraph{Particle initialisation} Particle initialisation is carried out by making a random draw from the respective parameter's prior distributions. 

\paragraph{Parameter constraints} For all parameters where positivity is a constraint (i.e. variance), the softplus transformation is applied with a clipping of $10^{-6}$, as is the default in GPFlow. Inference is then conducted on the unconstrained parameter, however, we report the re-transformed parameter i.e. the constrained representation.

\paragraph{Optimisation} We use the Adam optimiser \citep{kingma_adam:_2015} to optimise the set of \gls{SVGD} particles. The step-size parameter is started at 0.01 for the first 500 iterations, before being reduced to 0.005 and 0.001 for a further 500 iterations per step-size.

\paragraph{Data preprocessing} All the data used described in \secref{sec:background:data} is standardised to zero-mean and unit standard deviation.

\paragraph{Data availability}The LCM 2019 land cover data is available for download from the UKCEH Environmental Information Data Centre (EIDC). The dataset for GB is available at \url{https://doi.org/10.5285/f15289da-6424-4a5e-bd92-48c4d9c830cc} and the Northern Ireland dataset is available at \\ \url{https://doi.org/10.5285/2f711e25-8043-4a12-ab66-a52d4e649532}. The ERA5 ECMWF data is available for download from the Copernicus Climate Change Service (CS3) Climate Data Store (CDS) at \url{https://cds.climate.copernicus.eu/#!/search?text=ERA5&type=dataset}. Near real-time air quality data from the AURN, WAQN and SAQN networks are available through the Openair R package \citep{carslaw_openair_2012}.

\paragraph{Software} All code is written in Tensorflow \cite{abadi_tensorflow_2016} and extends the popular Gaussian process library GPFlow \citep{matthews_gpflow:_2017}. A full implementation of the SteinGP used in \secref{sec:analysis} can be found at \\ \href{https://github.com/thomaspinder/SteinGP}{https://github.com/thomaspinder/SteinGP} and corresponding experimental notebook will be publically released on submission.

\clearpage

\section{Demo Implementation}\label{app:demoImp}
\begin{python}
from steingp import SteinGPR, RBF, Median, SVGD
import tensorflow_probability.distributions as tfd
import numpy as np 
import gpflow

# Define some synthetic timestamps
X = np.random.uniform(-5, 5, 100).reshape(-1,1)
# Simulate a response plus some noise at each timestamp
y = np.sin(x) + np.random.normal(loc=0, scale=0.1, size=X.shape)

# Define model
kernel = gpflow.kernels.SquaredExponential()
model = SteinGPR((X, y), kernel)

# Place priors on hyperparameters
model.kernel.lengthscale.prior = tfd.Gamma(1.0, 1.0)
model.kernel.variance.prior = tfd.Gamma(2.0, 1.0)
model.likelihood.variance.prior = tfd.Gamma(2.0, 2.0)

# Fit 
opt = SVGD(RBF(bandwidth=Median()), n_particles=5)
opt.run(iterations = 1000)

# Predict
Xtest = np.linspace(-5, 5, 500).reshape(-1, 1)
theta = opt.get_particles()
posterior_samples = model.predict(Xtest, theta, n_samples=5)
\end{python}

\clearpage

\end{document}